\documentclass[11pt]{iopart}

\usepackage{iopams}

\usepackage[ansinew]{inputenc}
\usepackage{bm}
\usepackage[dvips]{graphicx}
\usepackage{color}

\newtheorem{theo}{Theorem}[section]
\newtheorem{lemma}{Lemma}[section]
\newtheorem{coro}{Corollary}[section]

\begin{document}

\title[On the invariant causal characterization]{On the invariant causal characterization of singularities in spherically symmetric spacetimes.}
\author{F. Fayos\thanks{Also at Laboratori de F\'{\i}sica Matem\`atica,
Societat Catalana de F\'{\i}sica, IEC, Barcelona.}  and R. Torres \\
Department of Applied Physics, UPC, Barcelona, Spain. \eads{\mailto{f.fayos@upc.edu}, \mailto{ramon.torres-herrera@upc.edu}}}

\begin{abstract}

The causal character of singularities is often studied in relation to the existence of naked singularities and the subsequent possible violation of the cosmic censorship conjecture. Generally one constructs a model in the framework of General Relativity described in some specific coordinates and finds an \emph{ad hoc} procedure to analyze the character of the singularity. In this article we show that the causal character of the zero-areal-radius ($R=0$) singularity in spherically symmetric models is related with some specific invariants. In this way, if some assumptions are satisfied, one can ascertain the causal character of the singularity algorithmically through the computation of these invariants and, therefore, independently of the coordinates used in the model.

\end{abstract}


\section{Introduction}
The causal character of a singularity has a well defined meaning within the theory of conformal boundaries \cite{Penrose63,RepSeno,charact}.
%
%
The knowledge of this causal character is fundamental since
whenever the spacetime possesses timelike or past null singularities
there are always null geodesics which are past incomplete.
If such a singularity could develop from a generic gravitational collapse in the framework of
General Relativity Theory, this would mean that the theory would lose its predictability. The question on whether General Relativity contains a built-in safety feature that precludes
this possibility was put forward by Penrose in 1969 \cite{Penrose} and gave rise to what is known as the \textit{cosmic censorship conjecture} (CCC).
Many counterexamples to the CCC have been proposed as well as many arguments in its favour (see, for example, \cite{charact,QCC} and references therein) so that the question of the CCC remains open.

In this article we will deal with probably the most interesting type of singularities in spherically symmetric spacetimes: The \textit{zero-areal-radius singularities},
i.e., given the \textit{areal radius} $R$ defined such that the area of a 2-sphere is $4\pi R^2$, we will be interested in singularities \textit{at} $R=0$.
In the current literature the study of the causal character of the $R=0$ singularities has been carried out for important particular solutions. In a few simple cases the singular conformal boundary has been obtained by using a conformal compactification
(see, for instance, \cite{H&E,Volovich,HisWiEar}),
while in most cases there is not an analytical compactification and, as an alternative method, the causal character of the singularities has been studied through the analysis of radial null geodesics around them (see, for example,
\cite{Kuroda,E&S,Chris,O&P}).
%
In addition to the analysis of particular cases, this last technique
allows some \emph{general} approaches for studying zero-areal-radius singularities. In particular, it has led to show \cite{Hayward} that the causal character of the singularity is related to the \textit{mass function} \cite{M&S}.
More general studies on the formation of \textit{naked} singularities in spherically symmetric spacetimes
along these lines
can be found in \cite{Lake}, \cite{Singh} and, by using ad hoc devised procedure, in \cite{GGMP}.
Furthermore, in \cite{charact}, by using the techniques of the qualitative behaviour of dynamic systems on the differential equations satisfied by the radial null geodesics we were able to present the most comprehensive scheme so far to try to find out their causal characterization taking into account, and analyzing, the possible limitations of the approach. However, this work was carried out in specific coordinates (the so called \textit{radiative coordinates}), so that its results were
restricted to models described in these coordinates.

Our aim in this article is to show that the causal character of the zero-areal-radius ($R=0$) singularity in spherically symmetric models is related with some specific invariants. Apart from being an interesting result from a theoretical point of view, this coordinate independent approach means that, if some assumptions are satisfied, one could find out the causal character of a model's singularity algorithmically through the computation of these invariants in arbitrary coordinates. In order to try to reach our goal we will base our approach in an analysis of the results in our previous article \cite{charact}. We will show that our previous results admit an interpretation and rewriting in terms of some invariants and we will analyze and explicitly state the limits for the applicability of our results coming from our specific approach. On the other hand, throughout the article we will use a geometrical approach requiring only the existence of a spacetime, but not the fulfillment of Einstein's equations.
Thus, we just try to discover the possibilities allowed by this geometrical approach which includes the classical as well as the semiclassical framework.


%

The paper has been divided as follows: In section \ref{Basis} we revise well-known properties of spherically symmetric spacetimes and of the radial null geodesics, but emphasizing the corresponding degrees of differentiability for each defined object (what will be an important aspect for the later development of the work). In section \ref{coordchange} we analyze the relationship between general coordinate systems and the coordinate system used in \cite{charact}. The different cases that appear when treating the causal character of singularities are treated from section \ref{m_neq0} on. In particular, section \ref{m_neq0} is devoted to the analysis of \textit{singularities with non-null mass function}. Section \ref{sec_m=0} deals with the preliminaries required to the study of \textit{singularities with null mass function}. Finally, sections \ref{sechyper}, \ref{secnonhyper} and \ref{secnoniso} analyze every \textit{null mass function} subcase in detail.

\section{Basis}\label{Basis}

Let us consider a simply connected open set $\mathcal V$ in a spherically symmetric spacetime and such that a part of its boundary consists of a $R=0$ interval.
The metric line element of an oriented spherically symmetric spacetime can be (and for practical purposes it is usually) written in the local chart endowed with coordinates $\{x^{\mu}\}=\{x^0,x^1,\theta,\varphi\}$ in the form \cite{Ple&Kra}
\begin{equation}
ds^2=g_{ij} dx^i dx^j+R^2\
d\Omega^2 , \label{mI}
\end{equation}
where $g_{ij}$ is an oriented two-dimensional Lorentzian metric ($i,j$=$0,1$), $R=R(x^0,x^1)$ and \mbox{$d\Omega^2\equiv
d\theta^2+\sin^2\theta d\varphi^2$}.


In the Lorentzian two-surface orthogonal to the 2-spheres
two nonvanishing null vector fields may be defined such that they are linearly independent at each point. If some differentiability requirements are satisfied in $\mathcal V$ the integral curves of the two null vector fields provide us with two families ($\mathcal{F}_1$ and $\mathcal{F}_2$) of
affinely parametrized null geodesics called the \textit{radial null geodesics}. Take, for example,
the geodesics belonging to the $\mathcal{F}_1$ family satisfying
\begin{displaymath}
\frac{d^2 x^\alpha}{d \ell^2}=-\Gamma^\alpha_{\beta\gamma} \frac{dx^\beta}{d \ell}\frac{dx^\gamma}{d \ell},
\end{displaymath}
where $\ell$ is their affine parameter. The theory of Ordinary Differential Equations \cite{Arnold} together with the definition of the Christoffel symbols $\Gamma^\alpha_{\beta\gamma}$ guarantees the existence and uniqueness of the affinelly parametrized geodesics provided that $g_{\alpha\beta}$ is at least $C^{2-}$ and $\det (g_{\alpha\beta})\neq 0$.
From now on we will guarantee the existence and uniqueness of affinely parametrized null geodesics by assuming that $g_{\alpha\beta}$ is $C^{n+1}$, where $n\geq 1-$\footnote{As usual $C^{i-}$ means $C^{(i-1)}$ and that the $(i-1)$th-order derivatives are locally Lipshitz. On the other hand, $C^{n+1}$ with $n=j-$ means $C^{(j+1)-}$.}, and that and $\det (g_{\alpha\beta})\neq 0$\footnote{Note that the last requirement, the non-degeneracy of the metric, is usually assumed in the definition of spacetime \cite{H&E}.}.

Under these assumptions the theory of ODEs also guarantees that the solution of the geodesic equation will be $C^{n+2}$.
Since the radial null tangent vector field to the $\mathcal{F}_1$ family $\vec{l}\equiv d/d\ell$ has an associated covector satisfying $l_{[\alpha ;\beta]}=0$, then it can be written as the differential of a $C^{n+2}$ function $u(x^0,x^1)$: $\partial_\alpha u\equiv l_\alpha$. The curves $u=$constant define the trajectories of the $\mathcal{F}_1$ family of null geodesics in $\mathcal V$.

Taking into account that the scalar invariant $R(x^0,x^1)$ is a $C^{n+1}$ function we can define $\chi(x^0,x^1)\equiv g^{\alpha\beta} \partial_\alpha R \partial_\beta R$ which is a scalar invariant $C^n$ function . This invariant is related to the $C^n$ invariant \textit{mass function} \cite{M&S,mass,mass2,mass3} through
\begin{equation}
m=\frac{R}{2}(1-\chi). \label{graven}
\end{equation}

In order to investigate the causal characterization of a $R=0$ interval we will consider the radial null geodesics around a point $p$ in this interval.
 As we shown in \cite{charact} this procedure
requires that the radial null geodesics from at least one family, say $\mathcal{F}_1$, \textit{reach} (or \textit{leave}) every point in the interval, what will be assumed in the next sections of this article.

On the other hand, let us comment that provided that a $R=0$ interval is not \textit{reached} (or \textit{left}) by radial null geodesics of any family then the causal characterization of the interval is straightforward since this
interval cannot be translated into a piecewise $C^1$ interval in the conformal boundary of the spacetime \cite{charact}. It can only be translated into a \emph{point} where the boundary is not a $C^1$ curve and, thus,
where there is not tangent vector properly defining its causal character.


\section{Analysis of a coordinate change}\label{coordchange}

As we mentioned in the introduction, the main goal of this article is to extend the coordinate dependent results presented in \cite{charact} so that they can be used independently of the coordinate system chosen to work with, i.e., to provide the causal characterization of $R=0$-singularities in a invariant manner. In order to do this, in this section we will deal with the connection between a general coordinate system of the type used for (\ref{mI}) and the coordinate system used in \cite{charact}.

\begin{lemma}\label{coocha}
Under the assumptions that the metric (\ref{mI}) is $C^{n+1}$, with $n\geq 1-$, and
$J\equiv u,_0 R,_1-u,_1 R,_0 \neq 0$
(or, equivalently, $du\wedge dR \neq 0$)
there exists a coordinate change $\{x^0,x^1\}\rightarrow \{u,R\}$, where $R$ is the areal coordinate and $u$ is a null coordinate, such that the metric (\ref{mI}) can be locally written as
\begin{equation}\label{metrad}
ds^2=-e^{4 \beta} \chi du^2+2\varepsilon e^{2 \beta} du dR+ R^2 d\Omega^2,
\end{equation}
where $\varepsilon^2=1$ and $\beta=\beta(u,R)$, $\chi=\chi(u,R)$ are $C^n$ functions.
\end{lemma}

This lemma is based in the fact that, since $u$ is $C^{n+2}$ and $R$ is $C^{n+1}$, there is a class $C^{n+1}$ map
$\Phi: (x^0,x^1) \rightarrow (u,R)$.
The inverse function theorem (see, for example, \cite{Boothby}) guarantees the existence of $C^{n+1}$ functions $f^0(u,R)$ and $f^1(u,R)$ such that $x^0=f^0$ and $x^1=f^1$ provided that the jacobian determinant $J$ is not null in $\mathcal V$. Along this work we will denote the open set $\Phi(\mathcal{V})$ by $\mathcal U$.

It follows that we can write the function $\chi$ as $\chi(u,R)=\chi(x^0=f^0 (u,R) , x^1=f^1(u,R))$, where the chain rule guarantees that $\chi(u,R)$ will also be at least                                                       a $C^n$ function in the variables $\{u,R\}$.

On the other hand, the condition $J\neq0$ or, equivalently, $du\wedge dR \neq 0$ implies that the vectors associated with these one-forms can not be parallel: $\vec{l}(R)\neq 0$\footnote{A sufficient (but not necessary) condition for $\vec{l}(R)\neq 0$ is $\chi= g^{\alpha\beta} \partial_\alpha R \partial_\beta R\neq 0$}. In this way, taking into account that $\vec{l}(R)$ will be $C^n$ ($n\geq 1-$) in $\mathcal V$ we can define  the invariant non-null constant
\begin{equation}
\varepsilon\equiv - \mbox{sign}(\vec{l}(R)).\label{varep}
\end{equation}
Note that it is invariant under future directed reparametrizations of $\vec l$.
If $\varepsilon=-1$ (or $+1$), the expansion \cite{H&E} of the null geodesics with tangent vector $\vec l$ is positive (negative, respectively) in every point of $\mathcal U$.

If we perform the coordinate change $\{x^0,x^1\}\rightarrow \{u,R\}$ then, due to the light-like character of the coordinate $u$, the metric of the spacetime (\ref{mI}) will take the form:
\begin{equation}
ds^2=-A du^2+2 B du dR+ R^2 d\Omega^2,
\end{equation}
where $A=A(u,R)$ and $B=B(u,R)$.
The general future directed and affinely parametrized null vector $\vec l$ and the future directed null vector $\vec k$ tangent to the $\mathcal F_2$ family satisfying $\vec l \cdot \vec k=-1$ can be written as
\begin{equation}\label{lk}
\vec l=-\frac{c}{B} \frac{\partial}{\partial R}  \ \ \ ; \ \ \
\vec k=\left( \frac{\partial}{\partial u}+ \frac{A}{2 B} \frac{\partial}{\partial R}\right) c^{-1},
\end{equation}
where $c=c(u)>0$ depends on the affine parameter chosen for $\vec l$. Clearly, sign$(\vec l(R))=-\varepsilon=-$sign$B$, so that we can rewrite $B$ as
\begin{equation}
B=\varepsilon e^{2\beta},\label{Bbeta}
\end{equation}
with $\beta=\beta(u,R)$.
If we state explicitly the relationship between the component $g_{uR}$ of the metric tensor in these new coordinates with regard to the old ones
\begin{equation}\label{beta}
\varepsilon e^{2\beta(u,R)}=
f^i,_u f^j,_R g_{ij},
\end{equation}
where, as usual, $g_{ij}$ should be understood as $g_{ij}(f^0(u,R),f^1(u,R))$, and taking into account that both $f^k(u,R)$ and $g_{ij}(x^0,x^1)$ are $C^{n+1}$ functions in their respective variables, then the chain rule theorem implies that $\beta(u,R)$ is at least a $C^n$ function.

On the other hand, if one evaluates the function $\chi$ with this form of the metric and uses (\ref{Bbeta}) one finds
\begin{equation}\label{Achi}
A=e^{4 \beta} \chi.
\end{equation}
Thus, taking into account (\ref{Bbeta}) and (\ref{Achi}),we obtain the required form for the metric (\ref{metrad}) with the degree of differentiability stated in the lemma. QED.

\begin{coro}\label{corm}
Under the assumptions in lemma \ref{coocha} the mass function $m(u,R)=R (1-\chi(u,R))/2$ is a $C^n$ function.
\end{coro}

Let us remark here that we do not need to find the explicit form (\ref{metrad}) since we want to work in the original coordinates $\{x^\mu\}$. In particular, the function $\beta$ can be written with the help of (\ref{beta}), and the appropriate labeling\footnote{So that ``$l_1$"$\neq 0$.}, as a function of $\{x^0,x^1\}$:
\begin{equation}\label{betax}
e^{2\beta(x^0,x^1)}=\varepsilon \frac{(l_0/l_1) g_{11} -g_{01}}{J}.
\end{equation}

\section{Invariants that define the causal character when $\lim_{x\rightarrow p} m(x)\neq 0$}\label{m_neq0}

In this section we will discuss the causal character of $R=0$ in a point $p=(p^0,p^1)$ such that $\lim_{x\rightarrow p} m(x)\neq 0$, where $x\equiv(x^0,x^1)$. This condition implies that there is a scalar curvature singularity at $p$ \cite{charact}, so that $p$ does not belong to the spacetime, but to its singular boundary. If one radial null geodesic of, say, the $\mathcal{F}_1$ family \textit{reaches} $p$ then we have the following
\begin{theo}\label{t_m_neq0}
In case the spacetime metric (\ref{mI}) is $C^{2-}$ and there is a radial null geodesic reaching (either toward its past or its future) a $R=0$-singularity at $p$ with a value of its affine parameter $l=l_0$ then:
\begin{itemize}
\item if $\lim_{l\rightarrow l_0} m(x(l))> 0$, there is a spacelike singularity at $p$,
\item if $\lim_{l\rightarrow l_0} m(x(l))< 0$, there is a timelike singularity at $p$.
\end{itemize}
\end{theo}

This result can be found in our previous article \cite{charact} where it was shown in specific coordinates. However, taking into account that $m$ is a scalar invariant, the proposition is true for other coordinate systems and it is, thus, an invariant result. Let us reiterate that the requirement of, at least, a $C^{2-}$ metric is a minimum assumption for the existence and uniqueness of radial null geodesics in the spacetime. The theorem only requires the existence of the \emph{directional} limit along the radial null geodesic. As a corollary we have the following more applicable result:
\begin{coro}\label{c_m_neq0}
In case the spacetime metric is $C^{2-}$ and $\lim_{x\rightarrow p} m(x)\neq 0$, where $p$ is such that $\lim_{x\rightarrow p} R(x)= 0$,  then the causal character of the $R=0$-singularity at $p$ is defined by the sign of the invariant mass function as the function approaches $p$:
\begin{itemize}
\item if $\lim_{x\rightarrow p} m(x)> 0$ then there is a spacelike singularity at $p$,
\item if $\lim_{x\rightarrow p} m(x)< 0$ then there is a timelike singularity at $p$.
\end{itemize}
\end{coro}

\section{Basis for the $\lim_{x\rightarrow p} m(x)= 0$ case} \label{sec_m=0}

This is the most involved case. It is known that if $\lim_{x\rightarrow p} m(x)= 0$ the causal character of $R=0$ at $p$ admits any possibility: spacelike, lightlike or timelike \cite{charact}. Therefore, the question is: are in this case other invariants which define the causal character of $R=0$ independently of the coordinate system used? In order to answer this, let us introduce in this section some new definitions and lemmas. For instance, in spherical symmetry there is an invariantly defined vector, the Kodama vector \cite{Kodama}, which is also known to possess very interesting properties (see, for instance, \cite{prop_Kodama} and references therein):
\begin{equation}
\vec{\xi}\equiv \epsilon_\bot^{\mu\nu} \frac{\partial R}{\partial x^\mu} \frac{\partial }{\partial x^\nu},
\end{equation}
where $\epsilon_\bot^{\mu\nu}$ denotes the volume form associated with the two-metric $g_{ij}$.
It satisfies $\vec{\xi}^2=-\chi$, so that if $\chi>0\ (\Leftrightarrow R>2 m)$ the orientation of $\epsilon_\bot^{\mu\nu}$ can be chosen in such a way that $\vec{\xi}$ is a future directed timelike vector.
Kodama's vector characterizes the spherically symmetric directions tangent to the hypersurfaces $R=$constant and provides and invariantly defined direction in which the area of the two-spheres remains constant \cite{SenoBeng}.
For the metric (\ref{metrad})
$\vec{\xi}$ takes the form:
\begin{equation}\label{KodRad}
\vec{\xi}= e^{-2 \beta} \partial_u,
\end{equation}
which satisfies $\vec l \cdot \vec \xi <0$ in the local chart.

\subsection{On the extension of $m(u,R)$ and $\beta(u,R)$}
We have seen that, if the assumptions in lemma \ref{coocha} are satisfied, the functions $m(u,R)$ and $\beta(u,R)$ are $C^n$ functions defined in an open set $\mathcal U$. We will now see that these functions can be extended beyond the open set provided that some conditions are fulfilled. In particular, we are mainly interested in an extension of the functions around $R=0$ that, while it has not any physical meaning, will allow us to apply the theory of the qualitative behaviour of dynamic systems in a open set centered in a point of $R=0$ \cite{charact}.
\begin{lemma}\label{extens}
Provided that there is a natural number $N$ such that the limits of the functions $m$ and $\beta$ and of their ith-order derivatives, for all $i\leq N$, as every point in the boundary of the open set is approached, exist (and are finite) then
the functions $m$ and $\beta$ admit a $C^N$ extension $\bar{m}$ and $\bar{\beta}$.
\end{lemma}
In order to show this it suffices to define the extended function and its derivatives for all points $p_b$ in the boundary of the open set $\mathcal U$ as
\begin{eqnarray*}
\frac{\partial^a\partial^b}{\partial u^a\partial R^b} \bar{m}(p_b)\equiv \lim_{x\rightarrow p_b} \frac{\partial^a\partial^b}{\partial u^a\partial R^b} m(u,R),\nonumber \\
\frac{\partial^a\partial^b}{\partial u^a\partial R^b} \bar{\beta}(p_b)\equiv \lim_{x\rightarrow p_b} \frac{\partial^a\partial^b}{\partial u^a\partial R^b} \beta(u,R)\nonumber
\end{eqnarray*}
for all integers $a\geq 0$, $b\geq 0$ such that $a+b\leq N$. In this way the lemma can be considered as a simple case of Whitney's extension theorem \cite{Whitney} and the existence of $C^N$ extensions $\bar{m}$ and $\bar{\beta}$
that coincide with $m$ and $\beta$ in $\mathcal U$
is guaranteed. We will denote the extended domain of definition by $\bar{\mathcal U}$.

On the other hand, we want to work with the original coordinates $\{x^0, x^1\}$ and infer from here the extendibility of the functions $m(u,R)$ and $\beta(u,R)$. We can do this by using the chain rule applied to the derivatives of $m(x^i(u,R))$ and $\beta(x^i(u,R))$ as the \emph{boundary points} are approached. For example, if one is looking for a $C^N$ extension, according to lemma \ref{extens} one needs, among others, the limit of $m,_u(u,R)$ in the boundary, but
\begin{displaymath}
\bar{m},_u(p_b)\equiv \lim_{x \rightarrow p_b} m,_u= \lim_{x\rightarrow p_b} m,_i f^i,_u.
\end{displaymath}
Taking into account that $f^i,_u$ can be written as a function of $l_j$ ($=u,_j$) and $R,_j$ ($j=0,1$), the existence of the limit is guaranteed if we require that the limits of $l_i(x^0,x^1)$, $R,_i(x^0,x^1)$,
and $m,_i(x^0,x^1)$
as every point in the boundary of the open set is approached exist (and are finite) and we also require that the limit of $J$ when approaching the same points is not zero.

This can be formalized and generalized for $C^N$ extensions similarly:
\begin{lemma}\label{extensCn}
If the limits
\begin{eqnarray*}
\lim_{x\rightarrow p_b} \frac{\partial^a\partial^b\ u}{\partial (x^0)^a\partial (x^1)^b} \ , \
\lim_{x\rightarrow p_b} \frac{\partial^a\partial^b\ R}{\partial (x^0)^a\partial (x^1)^b} ,\nonumber \\
\lim_{x\rightarrow p_b} \frac{\partial^a\partial^b\ m}{\partial (x^0)^a\partial (x^1)^b} \ ,\ 
\lim_{x\rightarrow p_b} \frac{\partial^a\partial^b\ \beta}{\partial (x^0)^a\partial (x^1)^b}\nonumber
\end{eqnarray*}
as every point in the boundary of the open set is approached exist (and are finite) for all integers $a\geq 0$, $b\geq 0$ such that $a+b\leq N$ and the limit of $J$ as the same points are approached exists and is not zero, then
the functions $m(u,R)$ and $\beta(u,R)$ admit a $C^N$ extension $\bar{m}(u,R)$ and $\bar{\beta}(u,R)$.
\end{lemma}

\subsection{On the differentiability of the singular conformal boundary}

In order to analyze the singular conformal boundary of a spherically symmetric spacetime it is possible to perform just the conformal compactification of the two-dimensional surface orthogonal to the \textit{2-spheres} retaining all the important information. This is so because, by means of a coordinate change, the induced Lorentzian metric or \textit{first fundamental form} of the two-dimensional surface can be brought into a conformally flat form $ds^2_{2d}= \Omega^2 (t,x) (-dt^2+dx^2)$ or, equivalently,
\begin{equation}\label{dobnul}
ds^2_{2d}= - \Omega^2 (u,v) du dv,
\end{equation}
where $\{u,v\}$ are lightlike coordinates ($u=t+x$, $v=t-x$).
In this way, it can be naturally embedded in an {\it unphysical} two-dimensional Minkowskian spacetime (see, for instance, \cite{RepSeno}).
Despite an \emph{analytic} conformal compactification  can only be found for certain simple particular cases some information can be extracted without fully following the procedure. For instance, with regard to the differentiability of the singular boundary we have the following
\begin{theo}\label{teoC3}
If $g_{ij}$ and $R$ are $C^{3}$ functions, $J\neq 0$ in $\mathcal V$ and
the assumptions in lemma \ref{extensCn} (for the existence of $C^2$ extensions) are satisfied and, in particular, they are satisfied in a connected open interval $\mathcal C$ of the singular boundary $R=0$ in which $\bar{m}\neq 0$,
then
$\mathcal C$ will be $C^{3}$ in the unphysical spacetime.
\end{theo}
In order to show this we will study the differentiability of $R=0$ in the unphysical spacetime by analyzing the coordinate change which takes the metric to the form (\ref{dobnul}).
To begin with, note that, under the assumptions in the theorem we can perform a first coordinate change from coordinates $\{x^0,x^1\}$ to $\{u,R\}$ such that $m(u,R)$ and $\beta(u,R)$ will be $C^{2}$ functions. That this is satisfied follows from the requirement of a $C^{3}$ metric in the theorem and lemma \ref{coocha} together with corollary \ref{corm}.
Furthermore, requiring the fulfillment of the assumptions in lemma \ref{extensCn} implies that there will be $C^2$ extensions $\bar{m}(u,R)$ and $\bar{\beta}(u,R)$.

We know from (\ref{metrad}) that the metric of the lorentzian surface can be written in coordinates $\{u,R\}$ as
\begin{equation}
ds^2_{2d}=-\frac{e^{4\beta}}{R} \sigma du,
\end{equation}
where $\sigma\equiv(R-2 m) du-2 \varepsilon R e^{-2\beta} dR$. In order to rewrite this in double-null coordinates $\{u,v\}$ we will look for an integrating factor $e^\Phi$ such that
\begin{equation}\label{dv}
e^\Phi \sigma =dv.
\end{equation}
The integrability condition for (\ref{dv}) takes the form of the following first-order linear inhomogeneous PDE in $\bar{\mathcal U}$:
\begin{equation}
a \Phi,_u+ b \Phi,_R=c,
\end{equation}
where the functions $a$, $b$ and $c$ are defined as
\begin{eqnarray*}
a&\equiv& 2 R e^{-2\bar{\beta}},\\
b&\equiv& \varepsilon (R-2 \bar{m}),\\
c&\equiv& 4 R \bar{\beta},_u e^{-2\bar{\beta}} -\varepsilon (1-2 \bar{m},_R).
\end{eqnarray*}
Our assumptions imply that $a$ and $b$ are $C^{2}$ functions, while $c$ is $C^{1}$.  We now choose the curve $\mathcal C$ (which can be described as $\{R=0,u=s\}$, where $s$ is a parameter defined in a connected open real interval) as the curve of initial conditions for the PDE and, in addition, we choose the initial condition on $\mathcal C$ to be $\Phi(u,R=0)=\psi(u)$, where $\psi(u)$ is an arbitrary $C^2$ function. Then the tangent vector to $\mathcal C$ ($\vec w=d/ds$) and the characteristic direction have distinct projections on the $u$,$R$-plane ($a w^R-b w^u=-2 \varepsilon \bar{m}(u, R=0)\neq 0$ in $\mathcal{C}$). This together with the fact that the PDE is nondegenerate ($a^2+b^2\neq 0$ in an open neighborhood of $\mathcal C$ since $\bar{m}(u,R=0)\neq 0$) implies that the initial value problem has one and only one solution \cite{C&H} which will be a $C^{2}$ function $\Phi(u,R)$. In this way (\ref{dv})
can now be used to evaluate the slope of the singular boundary $R(u,v)=0$ in the unphysical spacetime:
\begin{equation}
\frac{dv}{du}\rfloor_{R=0}=-2 e^{\Phi(u,R=0)} \bar{m}(u,R=0).
\end{equation}
Since there are only $C^{2}$ functions in the right hand side of this differential equation, its solution $v(u)$ will be a $C^{3}$ curve in the $u$,$v$-plane. QED

As a corollary, under these assumptions, $\mathcal C$ admits a tangent vector whose causal character determines the causal character of $R=0$. In other words, if $\bar m>0$ the slope of the curve is negative so that the curve in the unphysical spacetime (therefore, the singularity) will be spacelike. Likewise, if $\bar m<0$ the singularity will be timelike. This just reiterates the results in theorem \ref{t_m_neq0}, which were then shown under less restrictive assumptions. The so far untreated case $\bar m = 0$ in which the singularity can be timelike, lightlike or spacelike and in which the degree of differentiability of the singular boundary can be just $C^0$ under the differentiability assumptions on this subsection will be the subject of the rest of the article.

\section{Invariants if $\lim_{x\rightarrow p} m(x)= 0$ and
$\lim_{x\rightarrow p} \vec{\xi}(m)(x)\neq 0$} \label{sechyper}

\begin{figure}
\includegraphics[scale=0.7]{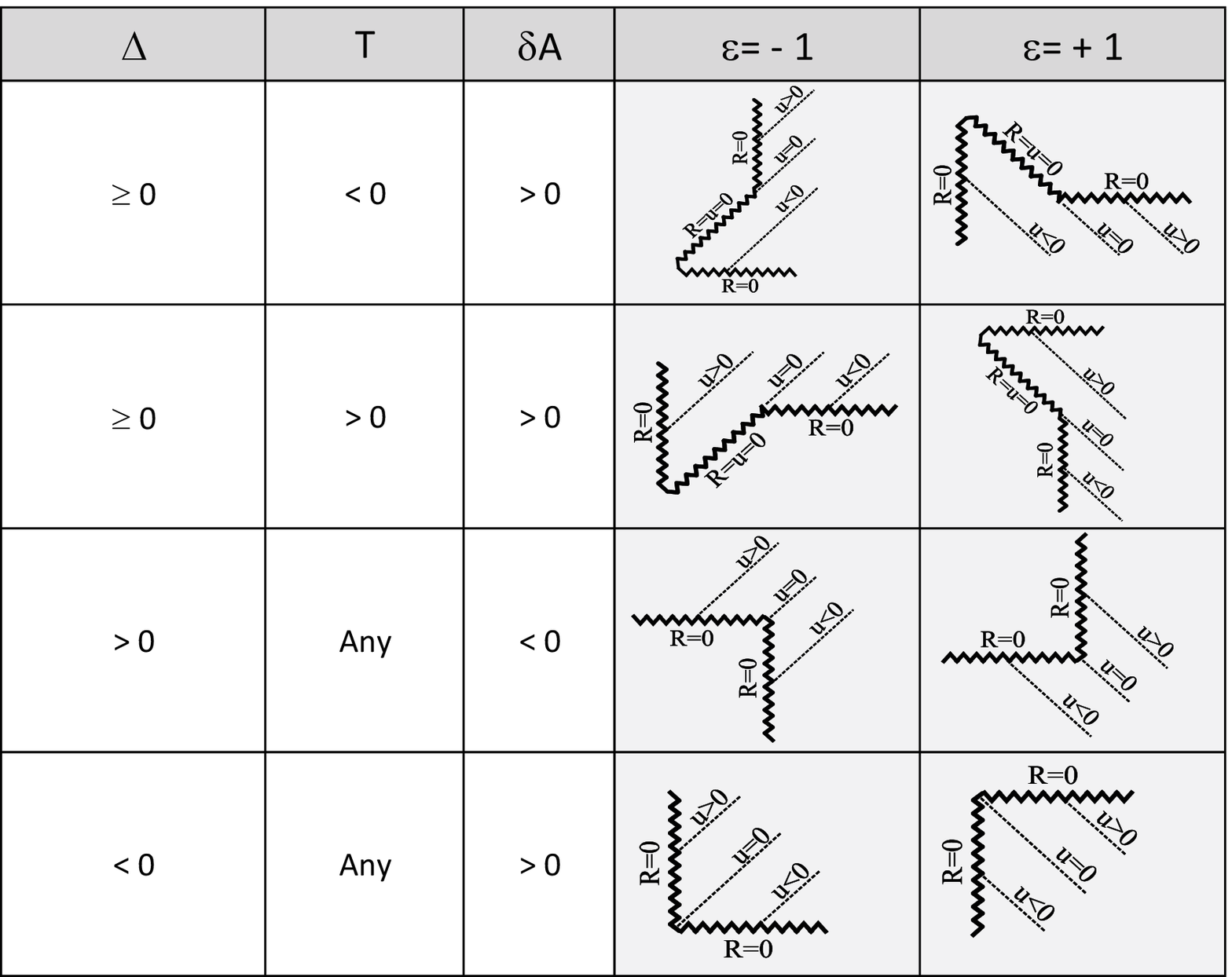}
\caption{\label{hyper} \emph{Sketched} characterization of the $R=0$ singularity when $\lim_{x\rightarrow p} m(x)= 0$ and
$\lim_{x\rightarrow p} \vec{\xi}(m)(x)=\mathcal{I}^1_b\neq 0$, where the $R=0$--singularity at $p$ is chosen to be reached by the radial null geodesic $u=0$.
As explained in the text, $\Delta =(1-2 \mathcal{J}^0_b)^2-16 \varepsilon \mathcal{I}^1_b$,
$T = \varepsilon  (1-2 \mathcal{J}^0_b)$ and $\delta A = -4 \varepsilon \mathcal{I}^1_b$. In this case we have $\bar{m}\rfloor_{R=0}\neq 0$ for $u\neq 0$. In this way, if the requirements in theorem \ref{teoC3} are satisfied the singular boundary should be $C^3$ for $u\neq 0$. (In these sketches we just draw straight lines for $u\neq 0$ instead of curves). However, in $u=0$, where $\bar{m}=0$ the singular boundary can be just $C^0$. Consider, for example, the first sketch in the $\varepsilon=-1$ column where a $R=0$-\emph{spacelike} singularity for $u<0$ must be abruptly followed by a lightlike singularity for $u=0$.}
\end{figure}

\begin{theo}\label{m0_mu}
In case $\lim_{x\rightarrow p} m(x)= 0$ and $\lim_{x\rightarrow p} \vec{\xi}(m)(x)\neq 0$, where $p$ is a point in the singular boundary $R=0$ reached or left by a single null geodesic of the $\mathcal{F}_1$ family, the causal character of $R=0$ around $p$ can be obtained through the invariants $\varepsilon$,
\begin{displaymath}
\mathcal{I}^1\equiv\vec{\xi}(m)=\xi^\alpha\partial_\alpha m\hspace{1cm} \mbox{and}\hspace{1cm} \mathcal{J}^0\equiv\frac{\vec{l}(m)}{\vec{l}(R)}=\frac{l^\alpha\partial_\alpha m}{l^\beta\partial_\beta R}.
\end{displaymath}
provided that $g_{ij}$ and $R$ are $C^{3-}$ functions in $\mathcal V$ and that
the assumptions in lemma \ref{extensCn} (for the existence of $C^1$ extensions) are satisfied.
Then the causal characterization around $p$ is inferred from $\varepsilon$,
\begin{displaymath}
\mathcal{I}^1_b\equiv\lim_{x\rightarrow p} \mathcal{I}^1 \hspace{1cm}\mbox{and}\hspace{1cm}
\mathcal{J}^0_b\equiv\lim_{x\rightarrow p} \mathcal{J}^0
\end{displaymath}
according to figure \ref{hyper}.
\end{theo}


Note that the invariant $\mathcal{J}^0$ is independent of the parametrization of $\vec{l}$.

In \cite{charact} (sec. 5) the causal characterization in this case was obtained in radiative coordinates
provided that $m(u,R)$ and $\beta(u,R)$ were $C^{2-}$ functions. That this requirement is satisfied follows from the requirement of a $C^{3-}$ metric in the theorem and lemma \ref{coocha} together with corollary \ref{corm}.
Another requirement was that there should be $C^1$ extensions $\bar{m}(u,R)$ and $\bar{\beta}(u,R)$,
hence our requirement that the assumptions on lemma \ref{extensCn} for the existence of $C^1$ extensions
should be satisfied.
On the other hand, it was shown that, in radiative coordinates, the causal characterization depends on\footnote[1]{Note that in \cite{charact} we chose $\bar{\beta}(u,0)=0$ for easiness through the use of a coordinate change $u'=u'(u)$. We do not perform this coordinate change here, what causes ``exp($-2 \bar{\beta}(u=0,R=0)$)" factors in this article when compared with the corresponding expressions in \cite{charact}. }
\begin{eqnarray*}
\Delta \equiv(1-2 \bar{m},_R (u=0,R=0))^2-16 \varepsilon \bar{m},_u(u=0,R=0) e^{-2 \bar{\beta}(u=0,R=0)},\\
\delta A \equiv -4 \varepsilon \bar{m},_u (u=0,R=0) e^{-2 \bar{\beta}(u=0,R=0)},\\
T\equiv \varepsilon  (1-2 \bar{m},_R (u=0,R=0)),
\end{eqnarray*}
where we have chosen to analyze the $R=0$-singularity around $u=0$ which describes the null geodesic reaching or leaving $p$ (what is always allowed through a redefinition of $u=u_{old}+$contant).

On the other hand, using the expression (\ref{lk}) for $\vec{l}$ and the Kodama vector (\ref{KodRad}) it is easy to verify that, in radiative coordinates, the invariant $\mathcal{I}^1$ can be written as $e^{-2\beta} m,_u$ while the invariant $\mathcal{J}^0$ is simply $m,_R$. Therefore, taking the limit $(u\rightarrow 0, R \rightarrow 0)$, we can rewrite the quantities $\Delta$, $\delta A$ and $T$ in an \emph{explicit invariant form} as
\begin{eqnarray*}
\Delta =(1-2 \mathcal{J}^0_b)^2-16 \varepsilon \mathcal{I}^1_b,\\
\delta A = -4 \varepsilon \mathcal{I}^1_b,\\
T = \varepsilon  (1-2 \mathcal{J}^0_b)
\end{eqnarray*}
Finally, in \cite{charact} it was shown that the causal character can be read from a table (figure \ref{hyper}) depending on these quantities.
In this way,
one first computes $m$, $\vec{\xi}$ and $\vec{l}$ in arbitrary coordinates in order to get from them
the invariants $\varepsilon$, $\mathcal{I}^1_b$ and $\mathcal{J}^0_b$.
Then, if the assumptions in the theorem are satisfied,
these invariants provide us with the causal characterization of the singularity for this case.

\section{Invariants if $\lim_{x\rightarrow p} m(x)= 0$ and
$\lim_{x\rightarrow p} \vec{\xi}(m)(x)= 0$}\label{secnonhyper}

\begin{figure}
\includegraphics[scale=0.7]{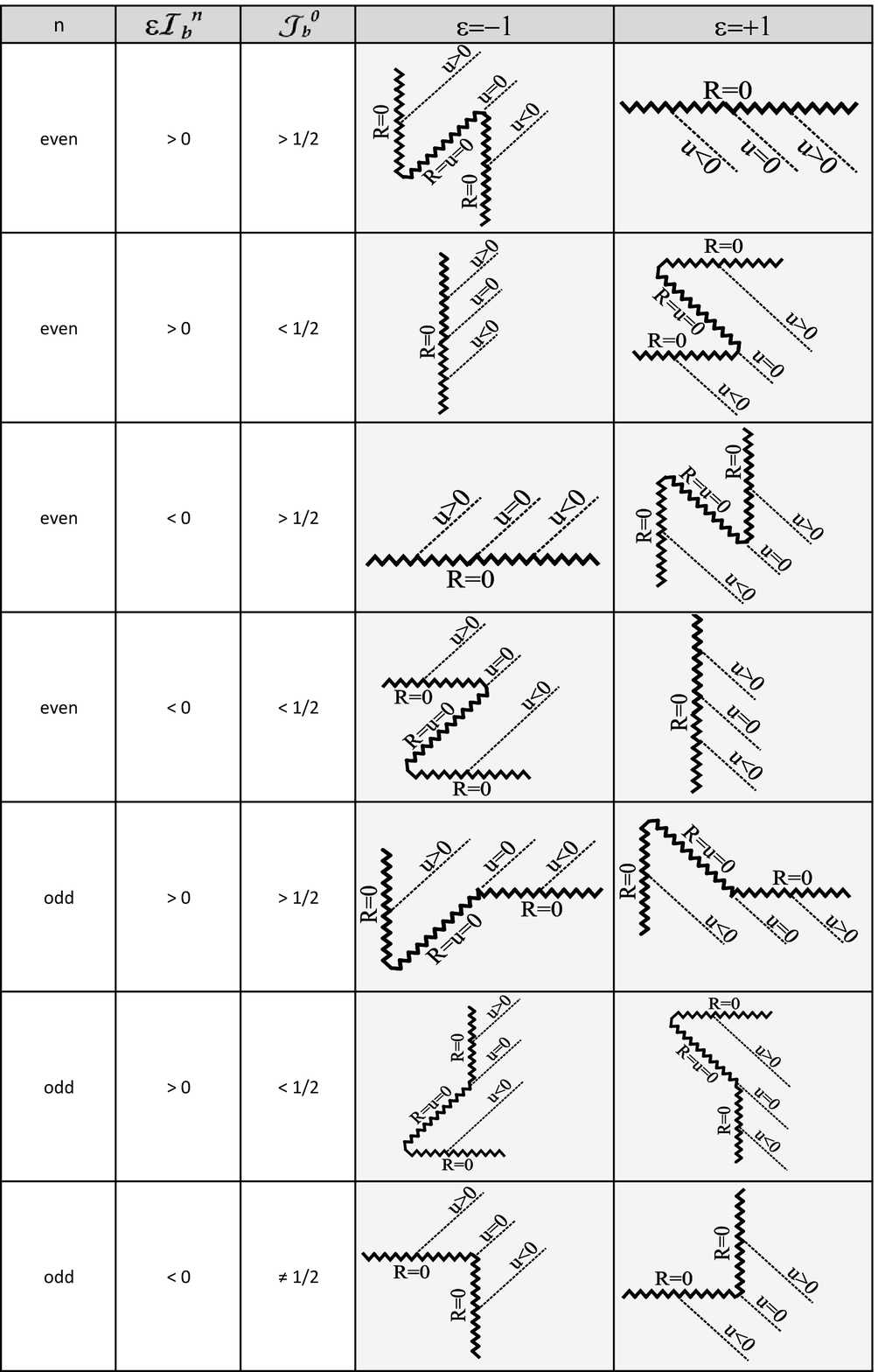}
\caption{\label{semihyper} Characterization of the singularity when $\lim_{x\rightarrow p} m(x)= 0$, $\lim_{x\rightarrow p} \vec{\xi}(m)= 0$ and $\lim_{x\rightarrow p} \vec{l}(m)/\vec{l}(R)\neq 1/2$.}
\end{figure}

\begin{theo}\label{m0_mu_mR}
In case $\lim_{x\rightarrow p} m(x)= 0$, $\lim_{x\rightarrow p} \vec{\xi}(m)= 0$ and $\lim_{x\rightarrow p} \vec{l}(m)/\vec{l}(R)\neq 1/2$, where $p$ is a point in the singular boundary $R=0$ reached or left by a single null geodesic of the $\mathcal{F}_1$ family, the causal character of $R=0$ is given by the invariants $\varepsilon$,
\begin{displaymath}
\mathcal{I}^n\equiv\vec{\xi}^n (m)=\stackrel{\mbox{n times}}{\overbrace{\xi^\alpha\partial_\alpha(\cdots(\xi^\omega\partial_\omega}}(m))\cdot)\hspace{1cm} \mbox{and}\hspace{1cm}
\mathcal{J}^0=\frac{\vec{l}(m)}{\vec{l}(R)}.
\end{displaymath}
provided that $g_{ij}$ and $R$ are $C^{n+1}$ functions in $\mathcal V$ ($n\geq 2$) and that
the assumptions in lemma \ref{extensCn} (for the existence of $C^n$ extensions) are satisfied. Then $\varepsilon$,
\begin{displaymath}
\mathcal{I}^n_b\equiv\lim_{x\rightarrow p} \mathcal{I}^n \hspace{1cm}\mbox{and}\hspace{1cm}
\mathcal{J}^0_b\equiv\lim_{x\rightarrow p} \mathcal{J}^0,
\end{displaymath}
where $\mathcal{I}^n_b\neq 0$ \footnote{Note that the definition of the case implies $\mathcal{I}^0_b=\mathcal{I}^1_b=0$.} whereas $\mathcal{I}^k_b=0$ for $0\leq k<n$ (i.e., we assume that there is a finite $n$ such that it is the lowest value satisfying $\mathcal{I}^n_b\neq 0$ ), provide us with the causal characterization of the singular boundary around $p$ according to figure \ref{semihyper}.
\end{theo}

In \cite{charact} (sec. 6) the causal characterization in this case was obtained in radiative coordinates
provided that $m(u,R)$ and $\beta(u,R)$ were $C^n$ ($n\geq 2$) functions admitting $C^n$ extensions. That this is satisfied follows, as in the previous theorem \ref{m0_mu}, first, from the requirement of a $C^{n+1}$ metric in this theorem and lemma \ref{coocha} together with corollary \ref{corm} and, second,
from the requirement that the assumptions in lemma \ref{extensCn} for the existence of $C^n$ extensions $\bar{m}(u,R)$ and $\bar{\beta}(u,R)$  are satisfied.
On the other hand, in \cite{charact} it was shown that, in radiative coordinates,
the causal characterization in this case
depends on\footnote{Here it has been taken into account that, when compared with the results in \cite{charact}, the calculation using $\bar{\beta}(u,R=0)\neq 0$ provides just an extra factor exp($-2 \bar{\beta}(0,0))>0$ to $\Delta_n$ which does not affect (\ref{deln}).}
\begin{eqnarray}
\mbox{sign}(\Delta_n)\equiv  \mbox{sign}\left(\varepsilon \frac{\partial^n \bar{m}}{\partial u^n}(u=0,R=0)\right) \hspace{.5cm}\mbox{and} \label{deln}\\
\frac{\partial \bar{m}}{\partial R}(u=0,R=0), \label{mR}
\end{eqnarray}

where $\partial^k \bar{m}/\partial u^k (u=0,R=0)=0$ for $0\leq k<n$ ($n\geq 2$).
But again these quantities are invariant. The second quantity (\ref{mR}) has already been shown to correspond with $\mathcal{J}^0_b$. With regard to the first one,
let us consider the simplest case $\mathcal{I}^2_b\neq 0$.
In radiative coordinates the invariant $\mathcal{I}^2$ is
\begin{displaymath}
\mathcal{I}^2=e^{-2\beta} [(e^{-2 \beta}),_u m,_u+e^{-2\beta} m,_{uu}].
\end{displaymath}
Since $\mathcal{I}^1_b=e^{-2 \bar{\beta}(0,0)} \bar{m},_u(0,0)=0$, in the limit as $(u\rightarrow 0,R\rightarrow 0)$ we will have
\begin{displaymath}
\mathcal{I}^2_b=e^{-4\bar{\beta}(0,0)} \bar{m},_{uu}(0,0).
\end{displaymath}
Likewise, in case $\mathcal{I}_b^k = 0$ for $0\leq k<n$ and $\mathcal{I}_b^n \neq 0$, we would have for $n\geq 2$
\begin{displaymath}\label{Inb}
\mathcal{I}^n_b=e^{-2 n\bar{\beta}(0,0)} \frac{\partial^n\bar{m}}{\partial u^n}(0,0).
\end{displaymath}
Therefore we can write the two quantities (\ref{deln}) and (\ref{mR}) in an explicit invariant form
\begin{displaymath}
\mbox{sign}(\Delta_n)\equiv  \mbox{sign} (\varepsilon \mathcal{I}^n_b) \hspace{.5cm}\mbox{and}\hspace{.5cm}
\mathcal{J}^0_b.
\end{displaymath}
Since these quantities define the causal character of $R=0$ in this case according to figure \ref{semihyper} \cite{charact}, then now the causal character has been determined invariantly.

\begin{figure}
\includegraphics[scale=0.7]{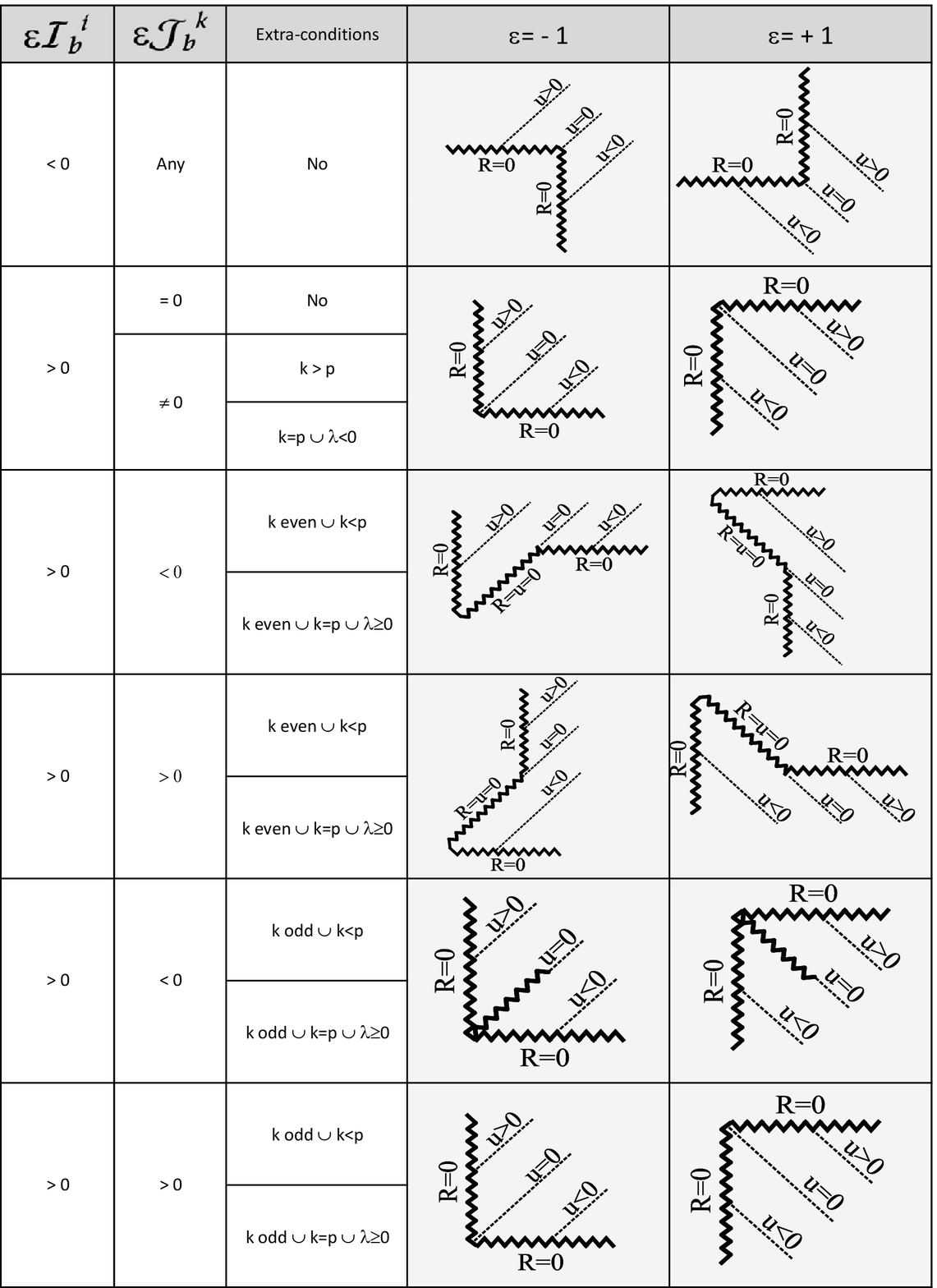}
\caption{\label{nilpotentodd} Characterization of the singularity when $\lim_{x\rightarrow p} m(x)= 0$, $\lim_{x\rightarrow p} \vec{\xi}(m)= 0$ and $\lim_{x\rightarrow p} \vec{l}(m)/\vec{l}(R)= 1/2$ and $i$ is odd. In the \textit{extra-conditions} we use $p\equiv(i-1)/2$ and $\lambda$ from (\ref{lambda}).}
\end{figure}

\begin{figure}
\includegraphics[scale=0.7]{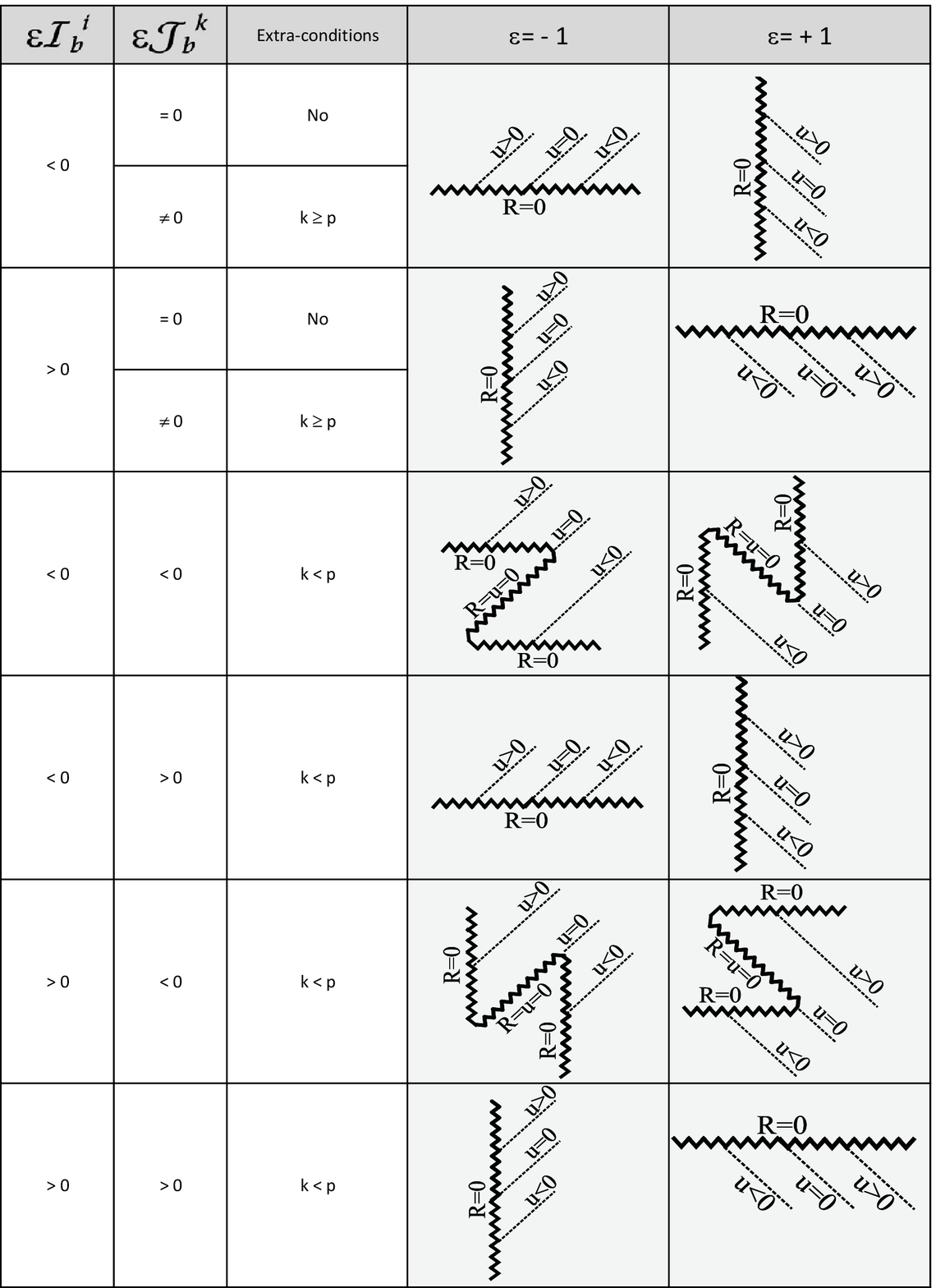}
\caption{\label{nilpotenteven} Characterization of the singularity when $\lim_{x\rightarrow p} m(x)= 0$, $\lim_{x\rightarrow p} \vec{\xi}(m)= 0$ and $\lim_{x\rightarrow p} \vec{l}(m)/\vec{l}(R)= 1/2$ and $i$ is even.  In the \textit{extra-conditions} we use $p\equiv i/2$.}
\end{figure}

\begin{theo}\label{m0_mu_mR0}
In case $\lim_{x\rightarrow p} m(x)= 0$, $\lim_{x\rightarrow p} \vec{\xi}(m)= 0$ and $\lim_{x\rightarrow p} \vec{l}(m)/\vec{l}(R)= 1/2$, where $p$ is a point in the singular boundary $R=0$ reached or left by a single null geodesic of the $\mathcal{F}_1$ family, the causal character of $R=0$ is given by the invariants $\varepsilon$,
\begin{displaymath}
\mathcal{I}^i\equiv\vec{\xi}^i (m)\hspace{1cm} \mbox{and}\hspace{1cm}
\mathcal{J}^k=\vec{\xi}^k\left(\frac{\vec{l}(m)}{\vec{l}(R)}\right).
\end{displaymath}
provided that $g_{ij}$ and $R$ are $C^{n+1}$ functions and that
the assumptions in lemma \ref{extensCn} (for the existence of $C^n$ extensions) are satisfied. Then  the causal characterization can be found according to figures \ref{nilpotentodd} and \ref{nilpotenteven} through the computation of
\begin{displaymath}
\mathcal{I}^i_b\equiv\lim_{x\rightarrow p} \mathcal{I}^i \hspace{1cm}\mbox{and}\hspace{1cm}
\mathcal{J}^k_b\equiv\lim_{x\rightarrow p} \mathcal{J}^k,
\end{displaymath}
where $\mathcal{I}^i_b\neq 0$ whereas $\mathcal{I}^j_b=0$ for $0\leq j<i$, $\mathcal{J}^k_b\neq 0$ whereas $\mathcal{J}^j_b=0$ for $1\leq j<k$ and $n\equiv$Max$\{i,k+1\}$ is supposed to exist ant to be finite.
\end{theo}

In \cite{charact} (sec. 6) the causal characterization in this case was obtained in radiative coordinates
provided that $m(u,R)$ and $\beta(u,R)$ were $C^n$ ($n\geq 2$, $n=$Max$\{i,k+1\}$, where $i$, $k$ are defined in (\ref{sak}) and (\ref{sbn}), respectively) functions admitting $C^n$ extensions. That this is satisfied follows, first, from the requirement of a $C^{n+1}$ metric in this theorem and lemma \ref{coocha} together with corollary \ref{corm} and, second,
from the requirement that the assumptions in lemma \ref{extensCn} for the existence of $C^n$ extensions $\bar{m}(u,R)$ and $\bar{\beta}(u,R)$ are satisfied.
According to \cite{charact} the causal characterization in this case and for radiative coordinates depends on sign($a_i$), sign($b_k$) and $\lambda\equiv b_k^2+2(i+1) a_i$, where\footnote{Note again that, unlike in \cite{charact}, factors exp($-2\bar{\beta}(0,0)$) appear since in this work we do not choose $\bar{\beta}(u,0)=0$. It has to be taken into account that if one now wishes to obtain $a_i$ and $b_k$ as in \cite{charact} there is a corresponding slight modification in the change of variables used there for this \textit{nilpotent case}: $x\equiv $exp($-2\bar{\beta}(0,0)) u/2$.}
\begin{eqnarray}
a_i\equiv  -\varepsilon e^{-2 i \bar{\beta}(0,0)}\frac{2^{i+1}}{i!} \frac{\partial^i \bar{m}}{\partial u^i}(u=0,R=0) \hspace{.5cm}\mbox{and}\label{sak}\\
b_k\equiv -\varepsilon e^{-2 k \bar{\beta}(0,0)}\frac{2^{k+1}}{k!}\frac{\partial^k \bar{m},_R}{\partial u^k}(u=0,R=0),\label{sbn}
\end{eqnarray}
with $\partial^j \bar{m}/\partial u^j (u=0,R=0)=0$ for $0\leq j<i$ and
$\partial^j \bar{m},_R/\partial u^j (u=0,R=0)=0$ for $1\leq j<k$.
But again these quantities are invariant. The first quantity sign($a_i$) has already been treated in the previous theorem where we showed that it corresponds to sign$(-\varepsilon \mathcal{I}^i_b)$. With regard to the second one sign($b_k$),
let us consider the simplest case $\mathcal{J}^1_b\neq 0$. In radiative coordinates the invariant $\mathcal{J}^1$ can be written as
\begin{displaymath}
\mathcal{J}^1=e^{-2\beta} m,_{R u}
\end{displaymath}
which in the limit as $(u\rightarrow 0,R\rightarrow 0)$ provide us with
\begin{displaymath}
\mathcal{J}^1_b=e^{-2\bar{\beta}(0,0)} \bar{m},_{R u}(0,0),
\end{displaymath}
which is clearly related to $b_1$ (\ref{sbn}). If $\mathcal{J}^1_b=0$ we should consider
\begin{displaymath}
\mathcal{J}^2=e^{-2\beta} [ (e^{-2 \beta}),_u m,_{R u}+ e^{-2 \beta} m,_{R u u}],
\end{displaymath}
which in the limit as $(u\rightarrow 0,R\rightarrow 0)$ provide us with
\begin{displaymath}
\mathcal{J}^2_b=e^{-4\bar{\beta}(0,0)} \bar{m},_{R u u}(0,0).
\end{displaymath}
Likewise, for $k\geq 2$ we will have
\begin{displaymath}
\mathcal{J}^k_b=e^{-2 k\bar{\beta}(0,0)} \frac{\partial^k\bar{m},_R}{\partial u^k}(0,0)
\end{displaymath}
to be compared with (\ref{sbn}).
In this way we can write the two quantities (\ref{sak}) and (\ref{sbn}) and, thus, their signs in an explicit invariant form
\begin{displaymath}
\mbox{sign}(a_i)\equiv  \mbox{sign} (-\varepsilon \mathcal{I}^i_b) \hspace{.5cm}\mbox{and}\hspace{.5cm}
\mbox{sign}(b_k)\equiv \mbox{sign}(-\varepsilon \mathcal{J}^k_b).
\end{displaymath}
Finally, taking into account its definition, $\lambda$ can also be written in an explicit invariant form
\begin{equation}\label{lambda}
\lambda=\frac{2^{2(k+1)}}{(k!)^2} (\mathcal{J}_b^k)^2-\varepsilon (i+1) \frac{2^{i+2}}{i!} \mathcal{I}_b^k.
\end{equation}
Since these quantities define the causal character of $R=0$ in this case according to figures \ref{nilpotentodd} and \ref{nilpotenteven}, then now the causal character has been determined invariantly for this case.

\section{Invariants if $\lim_{x\rightarrow p} m(x)= 0$ $\forall p \in \mathcal{C}$} \label{secnoniso}

\begin{theo}\label{noniso}
If $\lim_{x\rightarrow p} m(x)= 0$ $\forall p$ in a connected open interval $\mathcal{C}$ of $R=0$ then the causal character of $p$ is determined by the invariant
\begin{displaymath}
\mathcal{J}^0=\frac{\vec{l}(m)}{\vec{l}(R)}
\end{displaymath}
provided that $g_{ij}$ and $R$ are $C^{3-}$ functions in $\mathcal V$, that $m$ and $\beta$ are such that they admit $C^1$ extensions according to lemma \ref{extens} and that the limit
\begin{displaymath}
\mathcal{J}^0_b=\lim_{x\rightarrow p} \mathcal{J}^0
\end{displaymath}
exists (and is finite) for all $p \in \mathcal C$. Then the causal characterization around $p$ is inferred from
\begin{itemize}
\item If $\mathcal{J}^0_b<1/2$ then $R=0$ is timelike in this interval. (This case includes both the possibility of a singular or a \emph{regular} interval $\mathcal C$).
\item If $\mathcal{J}^0_b>1/2$ then $R=0$ is spacelike in this interval.
\item If $\mathcal{J}^0_b=1/2$ then $R=0$ is lightlike in this interval.
\end{itemize}
\end{theo}
In \cite{charact} (sec. 7) the causal characterization in this case was obtained in radiative coordinates
provided that $m(u,R)$ and $\beta(u,R)$ were $C^{2-}$ functions
admitting $C^1$ extensions $\bar{m}(u,R)$ and $\bar{\beta}(u,R)$. That these requirements are satisfied thanks to the assumptions in the theorem has already been shown for theorem \ref{m0_mu}.
On the other hand, in \cite{charact} it was shown that, in radiative coordinates, the causal characterization depends on $m,_R$ that we have shown can be written in an explicit invariant form as $\mathcal{J}^0$. In this way, the relationship between its value and the causal characterization follows directly from the results in \cite{charact} (sec. 7).

\section{Summary}

In this article we have shown that, provided some assumptions are satisfied, the causal character of the $R=0$--singularities in spherically symmetric spacetimes depends on some specific invariants. This allows to deduce the causal character of the singularity algorithmically. Basically, one starts with the knowledge of the areal radius $R$, the mass function $m$, a tangent vector field to the radial null geodesics $\vec l$ and the Kodama vector field $\vec{\xi}$. From here one should compute the invariants
\begin{displaymath}
\mathcal{I}^i=\vec{\xi}^i (m)\hspace{1cm} \mbox{and}\hspace{1cm}
\mathcal{J}^k=\vec{\xi}^k\left(\frac{\vec{l}(m)}{\vec{l}(R)}\right),
\end{displaymath}
where $i,k=0,1,2,...$ and the exact last value to be computed is determined by the values of the lowest order invariants as the singularity is approached ($\mathcal{I}^i_b$ and $\mathcal{J}^k_b$, respectively) in the following manner:
\begin{itemize}
\item If $\mathcal{I}^0_b=\bar{m}_b\neq 0$ this value suffices to characterize the singularity according to theorem \ref{t_m_neq0} provided the spacetime is at least $C^{2-}$ (i.e., if the existence and uniqueness of radial null geodesics is guaranteed).
\item If $\mathcal{I}^0_b=\bar{m}_b=0$ in an isolated point $p$ \textit{in} $R=0$ then different cases appear:
    \begin{itemize}
    \item If $\mathcal{I}^1_b\neq 0$ then we will also need $\mathcal{J}^0_b$. The causal characterization around the singular point can be inferred from figure \ref{hyper}, if the assumptions in theorem \ref{m0_mu} are satisfied.
    \item If $\mathcal{I}^1_b= 0$ and $\mathcal{J}^0_b\neq 1/2$ then one needs to compute $\mathcal{I}^i_b$, where we demand the existence of a finite $i\geq 2$ such that it is the lowest value satisfying $\mathcal{I}^i_b\neq 0$. The causal characterization around the singular point can then be inferred from figure \ref{semihyper}, if the assumptions in theorem \ref{m0_mu_mR} are satisfied.
    \item If $\mathcal{I}^1_b= 0$ and $\mathcal{J}^0_b= 1/2$ then one needs to compute $\mathcal{I}^i_b$, where we demand the existence of a finite $i\geq 2$ such that it is the lowest value satisfying $\mathcal{I}^i_b\neq 0$, and to compute $\mathcal{J}^k_b$, where we demand the existence of a finite $k$ such that it is the lowest value satisfying $\mathcal{J}^k_b\neq 0$. The causal characterization around the singular point can be inferred from figures \ref{nilpotentodd} and \ref{nilpotenteven}, if the assumptions in theorem \ref{m0_mu_mR0} are satisfied.
    \end{itemize}
\item If $\mathcal{I}^0_b=\bar{m}_b= 0$ in an open interval of $R=0$ then the causal characterization of the singular interval can be deduced from sign($\mathcal{J}^0_b-1/2$) according to theorem \ref{noniso}, if the assumptions in the theorem are satisfied.
\end{itemize}
Note that for every case some assumptions must be satisfied. These assumptions come mainly from the fact that our results are based on the use, in \cite{charact}, of the qualitative theory of dynamic systems to the differential equations satisfied by the radial null geodesics. The application of the appropriate theorems to the analysis of these differential equations requires some degree of differentiability for the functions $\bar m$ and $\bar \beta$. (More details on this issue can be found in \cite{charact}). Likewise, the reader can consult \cite{charact} for some applications of this technique to the study of the generation of naked singularities or black hole evaporation.

\section*{Acknowledgements}
We would like to thank J.M.M. Senovilla and Conan Wu for helpful discussions.
We would also like to acknowledge the \textit{Generalitat de Catalunya} (grant 2009SGR-00417) for financial support.

\end{document}